\documentclass[smallextended]{svjour3} 
\usepackage{epsfig}
\usepackage{amsmath}

\usepackage{graphicx}
\usepackage{dcolumn}
\usepackage{bm}
\usepackage{latexsym}
\usepackage{color}
\usepackage{cite}
\usepackage{hyperref}

\newcommand{\be}{\begin{equation}}
\newcommand{\ee}{\end{equation}}

\begin{document}

\title{Testing creation cold dark matter cosmology with the radiation temperature-redshift relation}

\author{Iuri P. R.  Baranov       \and
        Jos\'e F. Jesus \and
        Jos\'e A. S. Lima}


\institute{I. P. R. Baranov  \at
Departamento de Astronomia, Universidade de S\~ao Paulo, R. do Mat\~ao 1226, 05508-900, S\~ao Paulo, SP, Brazil \and
              Instituto Federal do Paran\'a, R. Felipe Tequinha 1400, 87703-536, Paranava\'i, PR, Brazil \\
               \emph{Instituto Federal de Educa\c{c}\~ao, Ci\^encia e Tecnologia da Bahia - Campus Sim\~oes Filho,
Via Universit\'aria s/n, Pitanguinha, 43700-000  - Sim\~oes Filho - Ba.} \\
	     \email{iuribaranov@gmail.com}   \\ 
           \and
           J. F. Jesus \at
              Universidade Estadual Paulista (Unesp), C\^ampus Experimental de Itapeva, R. Geraldo Alckmin 519, 18409-010, Itapeva, SP, Brazil \and Universidade Estadual Paulista (Unesp), Faculdade de Engenharia, Guaratinguet\'a, Av. Ariberto Pereira da Cunha 333, 12516-410, Guaratinguet\'a, SP, Brazil \\
              \email{jfjesus@itapeva.unesp.br}\\
              \and
              J. A. S. Lima \at
              Departamento de Astronomia, Universidade de S\~ao Paulo, R. do Mat\~ao 1226, 05508-900, S\~ao Paulo, SP, Brazil \\
              \email{jas.lima@iag.usp.br}}              

\date{Received: date / Accepted: date}

\maketitle


{Dark matter (stellar, interstellar, galactic, and cosmological)}
{Dark energy}

\def\zt{\mbox{$z_t$}}

\abstract{
The standard $\Lambda$CDM model can be mimicked at the background and perturbative levels (linear and non-linear) by a class of gravitationally  induced particle production cosmology dubbed CCDM cosmology. However, the radiation component in the CCDM model follows a slightly different temperature-redshift $T(z)$-law which depends on an extra parameter, $\nu_r$, describing the  subdominant photon production rate. Here we perform a statistical analysis based on a compilation of 36  recent  measurements of $T(z)$ at low and intermediate redshifts. The likelihood of the production rate in CCDM cosmologies is constrained by $\nu_r = 0.024^{+0.026}_{-0.024}$ ($1\sigma$ confidence level), thereby showing that  $\Lambda$CDM ($\nu_r=0$) is still compatible with the adopted data sample. Although being hardly differentiated in the dynamic sector (cosmic history and matter fluctuations), the so-called thermal sector (temperature law, abundances of thermal relics and CMB power spectrum) offers a clear possibility for crucial tests confronting $\Lambda$CDM and CCDM cosmologies.  }

\maketitle

\newpage


\vspace{0.1cm}
\section{Introduction}

It is widely believed that the present accelerating phase of the Universe is fueled by the vacuum energy density or cosmological constant present in the the top ten $\Lambda$CDM cosmology. However, the tiny value of the vacuum energy density associated with $\Lambda$ ($\rho_v=\Lambda/8\pi G)$ is plagued with the cosmic coincidence and cosmological constant problems \cite{Weinberg1989,Peebles2003,Padmanabhan2003}.  Such mysteries have  inspired many cosmologists to propose  alternative models based on rather different approaches \cite{Lima2004,Caldwell:2009ix}, among them: modified gravity \cite{mod_grav,Nojiri:2010wj,deMartino:2015zsa,Nojiri:2017ncd,Nojiri:2006ri,Capozziello:2011et}, quintessence \cite{RP1988,S2003,Carvalho2006,Demianski:2004qt}, time-evolving $\Lambda$ models \cite{Lam1,Lam2,Lam3,deMartino:2018cjh}, interaction between dark components \cite{Wang:2016lxa,Capozziello:2008it,Nunes:2016dlj,Jamil:2009eb,Chen:2008ft,Faraoni:2014vra} and gravitationally induced particle production or CCDM cosmology \cite{LJO2010,BL2010,BL2011,MP13,Komatsu2014,Pan,Waga2014,Waga2014a,LSC2016}. 

 
On the other hand,  it has been shown that a cosmology driven by gravitationally induced particle production
of all non-relativistic species existing in the present Universe mimics exactly the observed accelerating $\Lambda$CDM cosmology with just one dynamical free parameter \cite{LSC2016} (see also \cite{LJO2010} for creation of cold dark matter alone). This extended scenario also 
provides a natural reduction of the dark sector since the vacuum component is not needed
to accelerate the Universe.  The late time acceleration phase is obtained  with just one free parameter describing the effective production rate of the nonrelativistic components. The remarkable point here is that  such a cosmic scenario is fully degenerated with the $\Lambda$CDM model at the background, as well as when matter fluctuations are taken into account both in the linear and nonlinear 
levels (see also \cite{Waga2014,Waga2014a}). Different from the original CCDM cosmology \cite{LJO2010}, the associated creation process is also in agreement with the universality of the gravitational interaction in the sense that all nonrelativistic components are created and described by the specific creation rates.  Tests involving only the cosmic history or fluctuations of the matter component  cannot break the degeneracy presented by the $\Lambda$CDM and CCDM cosmologies.  

Naturally, the mentioned universality  implies that subdominant relativistic components (and relic neutrinos) are also ``adiabatically'' created by the evolving gravitational background.  This means that probes related to the physics of the thermodynamic sector cannot be discarded as a crucial test in breaking the degeneracy between $\Lambda$CDM and CCDM models both in the non-perturbative and perturbative levels. In particular, even considering that a blackbody spectrum with temperature $T_0=2.72548 \pm 0.00057$ is currently observed \cite{FixsenEtAl09,Aghanim:2018eyx}, the cosmic thermometers providing the temperature-law $T(z)$ must be slightly affected by the subdominant ``adiabatic'' creation of thermalized photons, as suggested long ago \cite{Lima:1995kd,L1,L2,L3} and more recently by the authors of references \cite{LJO2010,LSC2016} in their formulation  of the CCDM cosmology (see also \cite{Komatsu}).

 In this context, searching from departures in the $T(z)$ law may result in a degenerate splliting test. Cosmologies leading to $T(z)=T_0 (1+z)^{1-\alpha}$ have been tested using temperature data to constrain the $\alpha$ parameter \cite{deMartino:2012,deMartino:2015ema,deMartino:2016tbu,Avgoustidis:2015xhk,Luzzi:2009,Luzzi:2015via,Hurier:2014,Saro:2013fsr}. In similar way, by using a compilation of 36 $T(z)$ data at low and moderate redshifts, we confront here the predictions of $\Lambda$CDM and CCDM cosmologies. The former predicts the standard linear $T(z)$ law while in the latter, the temperature evolution is driven  by a nonlinear law  depending only on a single extra parameter which is constrained by these data over the interval $\nu_r = 0.024^{+0.026}_{-0.024}$  ($1\sigma$ confidence level). As we shall see, although being hardly differentiated in the dynamic sector  (cosmic history and matter fluctuations),  cosmic probes emerging from the thermal sector like the temperature-law, abundances of thermal relics and CMB power spectrum,  offer a clear possibility for crucial tests involving $\Lambda$CDM and CCDM cosmologies.

\section{CCDM Cosmology}

In this section, the formulation of the extended CCDM cosmology will be briefly reviewed.  The reader interested in more physical details may consult recent articles  \cite{LJO2010,BL2010,BL2011,MP13,Komatsu2014,Pan,Waga2014,Waga2014a,LSC2016}  and also the first papers  on the subject \cite{PRI89,LCW92,CLW92,LG92}. 

To begin  with, let us consider  a spatially flat FRW geometry whose time-evolving gravitational field is also a source of massive and massless particles. In the one-particle approach, the back reaction of the produced particles on the geometry is described by a creation pressure \cite{PRI89,LCW92,CLW92,LG92}. Here we consider a mixture as discussed in the extended approach where the Einstein field equations take the following form \cite{LSC2016}:

\begin{eqnarray}
&&8\pi G\sum_{i=1}^{N} \rho_i  = 3H^{2},\label{EE1}\\
&&8\pi G \sum_{i=1}^{N} (p_i + P_{ci})=-2\dot H - 3H^{2}, \label{EE2}\,
\end{eqnarray}
where  $H = \dot a/a$ is the Hubble parameter ($a(t)$ is the scale factor), an over-dot means time derivative with respect to the cosmic time  and $\rho_i$, $i=1,2,3\dots N$, denotes the energy density of a given component. In order to simplify matters we consider here only baryons,  cold dark matter (CDM) and the subdominant relic radiation.  The quantity $p_i$  denotes the equilibrium pressure defined by the usual equation of state (EoS) 

\begin{equation}\label{EoS}
 p_i = \omega_i \rho_i\,,\,\,\,(\omega_i=const \geq 0),
\end{equation}
while the negative creation pressure reads \cite{PRI89,CLW92,LG92}
\begin{equation}\label{pressaotermo}
 P_{ci} \equiv -(\rho_i + p_i)\frac{\Gamma_i}{3H}= - (1+ \omega_i) \frac{\rho_i \Gamma_i}{3H}, 
\end{equation}
where $\Gamma_i$ is the creation rate of each component and, for the last equality above, we have adopted Eq. (\ref{EoS}).   
In order to define a specific  matter creation  scenario,  a definite expression for  $\Gamma_i$ need to be assumed.  As shown in  Ref. \cite{LSC2016}, the one leading to a matter production model analogous to $\Lambda$CDM cosmology is defined by:

\begin{equation}\label{CR}
\frac{\Gamma_i}{3H} = {\alpha_i}\frac{\rho_{c0}}{\rho_{i}}.
\end{equation}
where $\alpha_i$  and $\rho_{c0}$ are, respectively, the creation rate of the i-th component and the present day value of the critical density.  The former should be determined by a more fundamental theory.  Note also that the presence of $\omega_i$ shows that $P_{ci}$ as given by (\ref{pressaotermo}) also depends on the nature of the created components. For the 3 basic  components assumed here (baryons, cold dark matter  and radiation) we have, respectively, the following negative creation pressures:

\begin{equation}\label{CRa}
P_{cb}= -{\alpha_{b}}{\rho_{c0}},\, \,\,\, P_{cdm} =  -{\alpha_{dm}}{\rho_{c0}},\,\,\,   P_{cr} = -\frac{4}{3}{\alpha_{r}}{\rho_{c0}}.
\end{equation}
Note that  each negative and constant creation pressure  is determined by an associated free parameter.  
All components are decoupled and the energy conservation law for each component reads: 

\begin{equation}\label{ECL}
\dot \rho_i + 3H(\rho_i + p_i + P_{ci})=0,
\end{equation}
and the total energy conservation law is contained in the field equations as should be expected (see \cite{LSC2016} for details). The solution of the above equation is given by:

\begin{equation}\label{CLI}
\rho_{i} = (\rho_{i0} - \alpha_i\rho_{c0})a^{-3(1 + \omega_i)} + \alpha_i\rho_{c0}\,, 
\end{equation}
where $\rho_{i0}$ is the present day energy density of the {\it i-th} component. 

Now, by neglecting the subdominant radiation contribution, it is readily seen that the total energy density  driving the cosmic dynamics reads:
\begin{equation}\label{CLIT}
\rho_T = \rho_{cdm} + \rho_b = \rho_{c0}\left[(1 - \alpha)a^{-3}+\alpha\right]\,, 
\end{equation}
where $\alpha = \alpha_{dm} + \alpha_b$. Note also that the Hubble parameter assumes the form:

\begin{equation}\label{H}
H^{2} = H_0^{2} \left[{(1 - \alpha})(1+z)^{3} + {\alpha} \right]\,.
\end{equation}

The above  equations (\ref{CLIT})-(\ref{H}) mimic exactly the $\Lambda$CDM expressions when the total clustering nonrelativistic matter density is identified by $\Omega_{eff}= 1-\alpha$. It is also worth noticing that the cosmic history and the perturbative expressions of the late CCDM cosmologies are sensitive only to the effective free parameter  $\alpha$  describing the total creation rate. This means that  the CCDM cosmology with creation of all nonrelativistic components  emulates the $\Lambda$CDM model with just one free parameter \cite{LSC2016}.

\section{CMB Temperature-law in CCDM Cosmology}

In this section we discuss a basic question of this paper, namely: How does the CMB temperature increase when we look back in time assuming the adopted CCDM cosmology? Such a question can be answered by using two different methods. The first approach is based on the nonequilibrum thermodynamics \cite{CLW92,LG92} while the second one (providing the same answer) comes from the associated kinetic formulation recently discussed \cite{LB14}. In order to emphasize the basic assumptions we next outline the thermodynamic derivation. Its kinetic counterpart is shortly presented in the Appendix A. 

\subsection{Temperature-Law and Thermodynamics}

The emergence of photons into the spacetime means that the balance equilibrium equations are modified. 
In particular, the particle flux  (${N^{\mu}_r}_{;\mu}=n_r\Gamma_r$) and entropy flux (${S^{\mu}_r}_{;\mu}=s_r\Gamma_r$) satisfy, respectively,   equations (A) and (B) below \cite{LB14}:  

\begin{equation}\label{BE}
(A)\,\,\,\, \dot n_r + 3Hn_r = n_r\Gamma_r, \, \, \,\,\,(B) \,\,\,\, \dot s_r + 3Hs_r = s_r\Gamma_r\,.
\end{equation}
where $n_r$, $s_r$ are, respectively, the concentration and the entropy density. We also recall that the Gibbs law reads \cite{DG}:

\begin{equation} 
n_rT_rd\sigma_r=d\rho - \frac{\rho_r + p_r}{n_r}dn_r,\,\label{entropyd}
\end{equation}
where $\sigma_r=s_r/n_r\equiv S_r/N_r$ is the especific entropy (per photon) [$S_r$ and $N_r$ are the entropy and number of photons per comoving volume]. Next, by taking  the pair ($T_r,n_r$) as independent thermodynamic variables, the above equation leads to:

\begin{equation}\label{TEMPa}
\frac{\dot T_r}{T_r}=\frac{1}{3}\frac{\dot n_r}{n_r}  + \frac{\dot \sigma_r}{(\frac{\partial \rho_r}{\partial n_r})_{T_r}}.
\end{equation}

By assuming  that the particles are generated by the evolving Universe in such a way that the specific entropy is constant (``adiabatic'' creation) it follows that:

\begin{equation}\label{S}
{\dot \sigma_r} = 0 \Leftrightarrow \frac{\dot S_r}{S_r} = \frac{\dot N_r}{N_r} =\Gamma_r,
\end{equation}
where in the last equality above the balance equations  presented in (\ref{BE}) were used. This condition means that $S_r=k_{B}N_r$ so that the irreversibilities leading to the entropy growth is due to the emergence of particles in the spacetime. By adopting the ``adiabatic'' constraint  and using again the equation for the particle concentration, the temperature law takes the general form  for an arbitrary $\Gamma_r$ expression: 
\begin{equation}\label{TEMPb}
\frac{\dot T_r}{T_r} =-\frac{\dot{a}}{a} + \frac{\Gamma_r}{3}. 
\end{equation}
In terms of the redshift parameter ($z=1/a-1$), the above equation is immediately integrated for an arbitrary creation rate \cite{LB14}:
\begin{equation}
T_r=T_0 (1+z)e^{-\int_0^z{\frac{\Gamma_r}{3H}\frac{dz'}{(1+z')}}},
\label{tempz}
\end{equation}
where $T_0$ is the temperature today. ``Adiabatic'' creation models are defined by specific creation rates $\Gamma_r$. In the case of CCDM cosmologies,  it is given by Eq. (\ref{CR}). Using the expression for the energy density  [see Eq. (\ref{CLI})], it is easy to show that:

\begin{equation}
\frac{\Gamma_r}{3H}=\alpha_r \frac{\rho_{c0}}{\rho_r}=\frac{\alpha_r}{\alpha_r + (\Omega_{r0}-\alpha_r)(1+z)^4},
\label{eq:}
\end{equation}
where $\Omega_{r0}=\rho_{r0}/\rho_{c0}$ is the present day radiation density parameter. By inserting this expression into (\ref{tempz}) a simple integration yields:
\begin{equation}\label{eq:Tznur}
T_r=T_0 (1+z)\left[1 + \nu_r \left(\frac{1}{(1+z)^4} - 1 \right) \right]^{1/4},
\end{equation}
where $\nu_r = \alpha_r/\Omega_{r0}$. Note that  for $\nu_r =0$ the radiation entropy is conserved and the standard linear temperature evolution, as predicted by the $\Lambda$CDM cosmology is recovered. 

The above temperature $T(z)$ law for CCDM cosmologies must be compared with the existing observations in order to constrain the free parameter $\nu_r$  or, equivalently, $\alpha_r = \nu_r {\Omega_{r0}}$. This is the main aim of next section.

\section {Data Sample and Statistical Analysis}

In Table \ref{TabTz}, we display the CMB $T(z)$ temperature data adopted here as provided by several authors in the literature 
\cite{FixsenEtAl09,LuzziEtAl09,SrianandEtAl08,NoterdaemeEtAl11,CuiEtAl05,GeEtAl97,SrianandEtAl00,HurierEtAl14}.

\begin{table}[h!]
\caption{The observed temperature--redshift relation. The 36 data points at low and intermediate redshifts are based on different observations, among them: the Sunyaev-Zeldovich effect, the rotational excitation of CO lines, the fine structure of carbon atoms, and X-ray data from galaxy clusters. By comparing  with the compilation presented by Komatsu and  Kimura \cite{Komatsu} we see that 11 measured values were drawn and 18 measured values were added.}
\label{TabTz}
\renewcommand{\tabcolsep}{1.5pc} 
\renewcommand{\arraystretch}{0.6} 
	\centering
		\begin{tabular}{c|c|c}
\hline
$z$ & $T$ (K) & Ref.\\
\hline
\hline
0	& 2.72548	$\pm$ 0.00057 & \cite{FixsenEtAl09} \\
$0.023$ &	$2.72 \pm 0.1$ & \cite{LuzziEtAl09}\\
0.152	& 2.9	$\pm$ 0.17 & \cite{LuzziEtAl09}\\
0.183	& 2.95 $\pm$ 0.27 & \cite{LuzziEtAl09}\\
0.2	& 2.74 $\pm$ 0.28 & \cite{LuzziEtAl09}\\
0.202	& 3.36 $\pm$ 0.2 & \cite{LuzziEtAl09}\\
0.216	& 3.85 $\pm$ 0.64 & \cite{LuzziEtAl09}\\
0.232	& 3.51	$\pm$ 0.25 & \cite{LuzziEtAl09}\\
0.252	& 3.39	$\pm$ 0.26 & \cite{LuzziEtAl09}\\
0.282	& 3.22	$\pm$ 0.26 & \cite{LuzziEtAl09}\\
0.291	& 4.05	$\pm$ 0.66 & \cite{LuzziEtAl09}\\
0.451	& 3.97	$\pm$ 0.19 & \cite{LuzziEtAl09}\\
0.546	& 3.69	$\pm$ 0.37 & \cite{LuzziEtAl09}\\
0.55	& 4.59	$\pm$ 0.36 & \cite{LuzziEtAl09}\\
2.418	& 9.15	$\pm$ 0.72 & \cite{SrianandEtAl08,NoterdaemeEtAl11}\\
1.777	& 7.2	$\pm$ 0.8 & \cite{CuiEtAl05}\\
1.973	& 7.9	$\pm$ 1 & \cite{GeEtAl97}\\
2.337	& 10	$\pm$ 4 & \cite{SrianandEtAl00}\\
0.037	& 2.888	$\pm$ 0.039 & \cite{HurierEtAl14}\\
0.072	& 2.931	$\pm$ 0.017 & \cite{HurierEtAl14}\\
0.125	& 3.059	$\pm$ 0.032 & \cite{HurierEtAl14}\\
0.171	& 3.197	$\pm$ 0.03 & \cite{HurierEtAl14}\\
0.22	& 3.288	$\pm$ 0.032 & \cite{HurierEtAl14}\\
0.273	& 3.416	$\pm$ 0.038 & \cite{HurierEtAl14}\\
0.322	& 3.562	$\pm$ 0.05 & \cite{HurierEtAl14}\\
0.377	& 3.717	$\pm$ 0.063 & \cite{HurierEtAl14}\\
0.428	& 3.971	$\pm$ 0.071 & \cite{HurierEtAl14}\\
0.471	& 3.943	$\pm$ 0.112 & \cite{HurierEtAl14}\\
0.525	& 4.38	$\pm$ 0.119 & \cite{HurierEtAl14}\\
0.565	& 4.075	$\pm$ 0.156 & \cite{HurierEtAl14}\\
0.618	& 4.404	$\pm$ 0.194 & \cite{HurierEtAl14}\\
0.676	& 4.779	$\pm$ 0.278 & \cite{HurierEtAl14}\\
0.718	& 4.933	$\pm$ 0.371 & \cite{HurierEtAl14}\\
0.777	& 4.515	$\pm$ 0.621 & \cite{HurierEtAl14}\\
0.87	& 5.356	$\pm$ 0.617 & \cite{HurierEtAl14}\\
0.972	& 5.813	$\pm$ 1.025 & \cite{HurierEtAl14}\\			
\hline
\hline
		\end{tabular}\\
\end{table}

  We notice that the $T_0\equiv T(z=0)$ data comes from the CMB blackbody radiation spectrum as estimated from \cite{FixsenEtAl09}, where he has recalibrated the FIRAS data from COBE \cite{MatherEtAl99} with the WMAP data \cite{HinshawEtAl09}. The 13 low redshift ($z=0.023-0.55$) $T(z)$ data are from \cite{LuzziEtAl09}, taken from multi-frequency measurements of the Sunyaev-Zeldovich (S-Z) effect toward 13 clusters (A1656, A2204, A1689, A520, A2163, A773, A2390, A1835, A697, ZW3146, RXJ1347, CL0016+16 and MS0451-0305 by order of redshift, shown on Table \ref{TabTz}). A value at high redshift comes from a damped Lyman-$\alpha$ (DLA) system \cite{SrianandEtAl08,NoterdaemeEtAl11}, from the CO excitation temperature. It was the first measurement of $T(z)$ using a molecular transition at high redshift. An additional data point  is also associated to a DLA system, Q1331+170 \cite{CuiEtAl05} which was found by applying the inferred physical conditions to the observed C I fine structure excitation. This same process has furnished a $T(z)$ value from QSO 0013-004 \cite{GeEtAl97}. Srianand {\it et al.} \cite{SrianandEtAl00} reported a CMB high-$z$ temperature from first and second fine structure levels of neutral carbon atoms in an isolated remote gas cloud. Finally, Hurier {\it et al.} \cite{HurierEtAl14} reported 18 $T(z)$ data by using thermal Sunyaev-Zeldovich (tSZ) data from the Planck satellite.

\begin{figure}[t!]
\centerline{\epsfig{figure=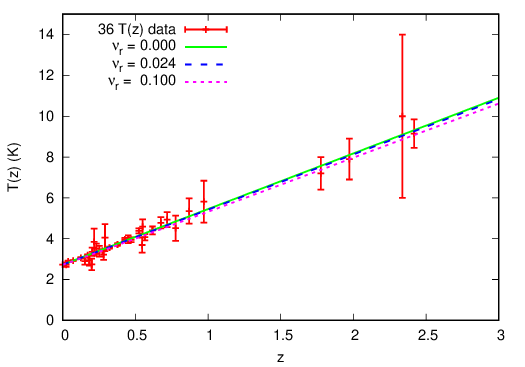,width=0.9\textwidth}\hskip 0.05in}
\caption{$T(z)$ data from many references, as listed on Table \ref{TabTz} and the prediction of some models as depicted in the figure. Note that the green solid line ($\nu_r=0$) corresponds to the standard $\Lambda$CDM.}
\label{fig1}
\end{figure}

Many of these data had already been compiled by Komatsu and Kimura \cite{Komatsu}. However, differently from the quoted authors,  5 asymmetric uncertainty data were not considered  in order to avoid a result dependent over the symmetrization method choice. In addition, the 6 $T(z)$ data deduced from Planck tSZ \cite{MartinoEtAl15} were not also used in order to prevent data duplicity.

 In Fig. \ref{fig1}, we display all data along with the predicted $T(z)$ curves obtained from (\ref{eq:Tznur}) for some selected values of $\nu_r$.  For ``adiabatic'' particle production, the second law of thermodynamics restricts $\nu_r$ to positive values \cite{PRI89}. In the present approach this can be seen directly from Eq. (\ref{S}).

\begin{figure}[t!]
\vspace{-.15in}
\centerline{\epsfig{figure=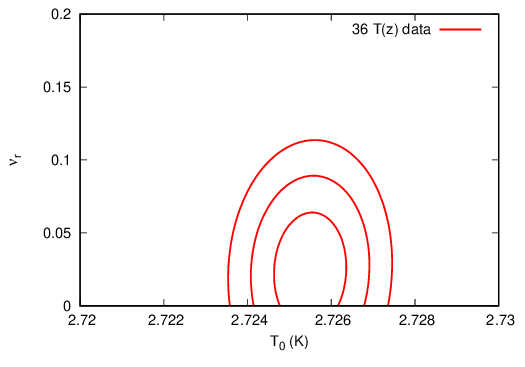,width=0.9\textwidth}
\hskip 0.05in}
\caption{Confidence contours on the  $(T_0,\nu_r)$ plane for the CCDM cosmology as inferred from 36 $T(z)$ data at low and intermediate redshifts (see Table \ref{TabTz} and Fig. \ref{fig1}).  The best fit values are located at $(T_0,\nu_r) \equiv (2.72549,0.024)$.  }
\label{fig2}
\end{figure}

With this data at hand, we performed a statistical analysis based on a $\chi^2$ procedure, by minimizing
\begin{equation}
 \chi^2=\sum_{i=1}^{36}\left(\frac{T(z_i)-T_i}{\sigma_{Ti}}\right)^2\,
\end{equation}
where $T(z)$ is the modeled temperature dependence (\ref{eq:Tznur}), $T_i$ is the observed temperature at redshift $z_i$ and $\sigma_{Ti}$ are their uncertainties. In addition, although considering that $T_0$ is nicely determined by the COBE+WMAP analysis, we choose to leave it as a free parameter, in order to check the consistence of the model. 

 In Fig. \ref{fig2}, we present the confidence contours on the  $(T_0,\nu_r,)$ plane. From the figure we see that the $(T_0,\nu_r)$ plane is nicely constrained by this analysis. The results obtained from the joint analysis are: $\nu_r=0.024^{+0.040 +0.065 +0.090}_{-0.024 -0.024 -0.024}$ and $T_0 = 2.72549^{+0.00087 +0.0014 +0.0020}_{-0.00086 -0.0014 -0.0020} $ K
 at 68.3\%, 95.4\% and 99.7\%, respectively. We also have found the minimal value, $\chi^2_{min} = 28.183$,  and the reduced $\chi^2_{red} = 0.829$.


\begin{figure}[t!]
\vspace{-.15in}
\centerline{\epsfig{figure=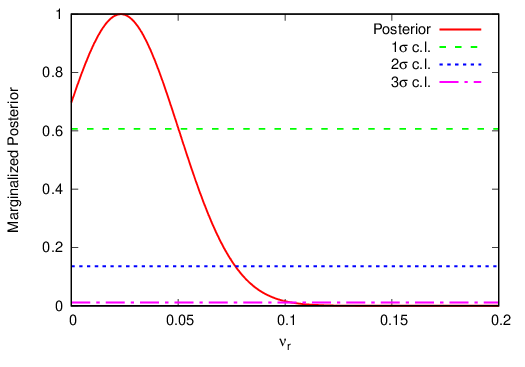,width=0.9\textwidth}
\hskip 0.05in}
\caption{The likelihood of the production rate free parameter $\nu_r$ in the CCDM cosmology. It is peaked for a positive value (creation) and is  constrained by $\nu_r = 0.024^{+0.026}_{-0.024}$ at 1$\sigma$ confidence level. Note that for this data sample the $\Lambda$CDM model ($\nu_r=0$) is compatible at 1 $\sigma$.}
\label{fig3}
\end{figure}

Next, in order to focus on the $\nu_r$ constraint, we marginalized the  posterior $\pi(\nu_r,T_0)\mathcal{L}(\mathrm{data},\nu_r,T_0)$ over $T_0$, where $\pi$ is the prior, which we assume to be flat and $\mathcal{L}$ is the likelihood $\mathcal{L}=Ne^{-\frac{\chi^2}{2}}$, where $N$ is a normalization constant.  As we assume the prior to be flat over a large range of the parameters ($\nu_r\in[0,10]$, $T_0\in[0,10]$)\footnote{ Another possible uninformative prior would be Jeffreys prior. However, this prior diverges for $\nu_r\rightarrow0$ and the number of available data is enough for the results to be weakly dependent on the choice of uninformative priors \cite{Jaynes03}, so we use only flat prior.}, to integrate over the posterior is equivalent to integrate over the likelihood (up to a multiplying constant). We find a ``marginalized'' $\tilde{\chi}^2$ from $\mathcal{\tilde{L}}(\nu_r)\equiv\int_{0}^{\infty}\mathcal{L}(\nu_r,T_0)dT_0\approx\int_{-\infty}^{\infty}\mathcal{L}(\nu_r,T_0)dT_0$. The last approximation is due to the fact that the likelihood is negligible in the region $T_0<0$ thanks to the available data. The last integral can be done analytically, as $T(z)$ is linearly dependent over $T_0$. We find
\begin{equation}
 \tilde{\chi}^2=-2\ln\left(\frac{\mathcal{\tilde{L}}}{N}\right)=S_{TT}-\frac{S_{fT}^2}{S_{ff}}
\end{equation}
where we have defined $f(z)\equiv\frac{T(z)}{T_0}$ from Eq. (\ref{eq:Tznur}) and
\begin{equation}
 S_{ff}\equiv\sum_{i=1}^{36}\frac{f(z_i)^2}{\sigma_{Ti}^2},\, S_{fT}\equiv\sum_{i=1}^{36}\frac{f(z_i)T_i}{\sigma_{Ti}^2},\, S_{TT}\equiv\sum_{i=1}^{36}\frac{T_i^2}{\sigma_{Ti}^2}.
\end{equation}

In Fig. \ref{fig3} we present the likelihood of the production rate $\nu_r$. We have found, from this analysis, $\nu_r = 0.024^{+0.026 +0.053 +0.078}_{-0.024 -0.024 -0.024}$, at 68.3\%, 95.4\% and 99.7\%, respectively. Note that for the central value of $\nu_r$ the $\alpha_r$ parameter modulating the creation rate, $\alpha_r = \nu_r\Omega_{r0} \sim 10^{-7}$.

\section{Conclusion}

 As briefly reviewed in the introduction, the $\Lambda$CDM model can be mimicked at the background and perturbative levels (linear and non-linear) by a class of gravitationally  induced particle production cosmology dubbed CCDM model \cite{LJO2010,Waga2014a,LSC2016}.

In this paper, we have discussed  a new probe to CCDM cosmology provided by the so-called  thermal sector, namely: the temperature-redshift $T(z)$-law. The predictions of CCDM cosmology were analyzed and we have demonstrated that this reduction of the dark sector with photon production passes this simple temperature test. 
We have used the available $T(z)$ data to constrain CCDM free parameters (see Figures \ref{fig1}, \ref{fig2} and \ref{fig3}). The derived nonlinear $T(z)$ expression in the present ``adiabatic'' context also preserves the blackbody form in the course of the expansion  (see \cite{L1,L3}), but  (\ref{eq:Tznur}) differs significantly from the linear law predicted by the $\Lambda$CDM model. The more general expression is modulated by a phenomenological free parameter ($\nu_r$)  associated to the  gravitationally induced photon production rate.

In order to constrain  $\nu_r$ we have carried out a statistical analysis based on 36 recent measurements of $T(z)$ at low and moderate redshifts. The posterior distribution of the production rate in CCDM cosmologies constrains its value to $\nu_r = 0.024^{+0.026}_{-0.024}$ ($1\sigma$ confidence level) thereby showing that  $\Lambda$CDM ($\nu_r=0$)  is also compatible (at $1\sigma$) with the adopted data sample (see Fig. \ref{fig3}). Due to the ``adiabatic'' photon production, our results suggest (for redshifts of the order of a few) that the CMB temperature in the context of CCDM models may be slightly lower than the value predicted by the $\Lambda$CDM cosmology. However, accurate data at low and moderate $z$, as well as another cosmological tests, such as the calculation of the angular power spectrum of the CMB temperature anisotropies, for example,  are needed in order to have a more definite conclusion.

Finally, by taking into account the results obtained here and the fact that measurements of $H_0$ at low and high redshifts are now endowed with a  $3\sigma$-level discrepancy (Supernovae-CMB tension \cite{Riess2016}),  we believe that  probes in the thermal sector which are responsible for the success of the fitting $\Lambda$CDM model should be carefully reinvestigated. In principle,  even conserved quintessence models may be ruled out by cosmic probes in the thermal sector, mainly when confronted with a model allowing a mild gravitationally induced production of photons. The present results reinforce the idea that the present observed $\Lambda$CDM description may be an effective CCDM cosmology. However, we are not advocating here that a crucial test has been obtained since both models are compatible with the unperturbed thermal history.


\section{Appendix A - Kinetic Theory and Temperature Law}

In this appendix, we use the extended Boltzmann equation for gravitationally particle production, proposed in previous works \cite{LB14,BL15}, to get the temperature evolution of the relic radiation. 

In a multi-fluid approach, each component has its own equation with its corresponding production rate. The extended Boltzmann equation describing this gravitational, non-collisional (each component evolves freely from the others), process is
\begin{equation}\label{boltz1} \tag{A1}
  \frac{\partial f_i}{\partial t}-H\left(1-\frac{\Gamma_i}{3H} \right)p\frac{\partial f_i}{\partial p}=0,
\end{equation}
where $f_i$ and $\Gamma_i$ are, respectively, the distribution function and the production rate for the $i$-th component.

For a relativistic quantum gas, the distribution function is \cite{landau}
\begin{equation} \tag{A2}
f=\frac{1}{e^{-\Theta+\beta E}+\epsilon},
\label{dist}
\end{equation}
where $\Theta$ is the relativistic chemical potential, $\beta = 1/T$, where $T$ is the temperature, and $\epsilon=\pm 1$ counts for different quantum statistics.
 
In the case of a relativistic bosonic gas with creation rate $\Gamma_r$ (CMB radiation), by inserting (\ref{dist}) into (\ref{boltz1}) we obtain (see also \cite{LB14}): 
\begin{equation}\tag{A3}
\frac{\dot \Theta}{\dot \beta}=E\left[1- H \frac{\beta}{\dot \beta}\left(1-\frac{\Gamma_r}{3H} \right)\right],
\label{CMBsol1}
\end{equation}
which has the solution $\dot \Theta=0$ and
\begin{equation}\tag{A4}
\frac{\dot T_r}{T_r}=-\frac{\dot a}{a}+\frac{\Gamma_r}{3},
\label{CMBsol2}
\end{equation}
where $a$ is the scale factor and $\Gamma_r$ stands for the radiation production rate. The above expression is the same as Eq. (\ref{TEMPb}) which was deduced based on the thermodynamic approach. In the limit $\frac{\Gamma_r}{3H}\ll 1$, equations (\ref{CMBsol1}) and (\ref{CMBsol2}) are reduced to the usual ones  (see equations (3.70) and (3.71) in \cite{bernstein}).

\vspace{1.0cm}
{\bf Acknowledgments:} The authors are partially supported by CNPq, FAPESP and CAPES
(LLAMA project, INCT-A and PROCAD2013 projects). JFJ acknowledges financial support from FAPESP, Process n$^\mathrm{o}$ 2017/05859-0, Funda\c{c}\~ao de Amparo \`a Pesquisa do Estado de S\~ao Paulo (FAPESP).


\end{document}